\documentclass[12pt]{article}
\usepackage{epsf,latexsym}
\usepackage{amsmath,amsfonts,amssymb,amsthm,cite}

\newtheorem{thm}{Theorem}[section]

\theoremstyle{definition}                    
\newtheorem{defn}[thm]{Definition}
\theoremstyle{remark}

\numberwithin{equation}{section}             

\usepackage[ps,dvips,arc,frame]{xy}

\epsfverbosetrue
\newcommand{\Mf}{\hbox{$^f$\hspace{-0.2cm} {\it M}}}

\newcommand{\ddf}{\hbox{$^f$\hspace{-0.15cm} $\mathcal{D}$}}

\newcommand{\ddfm}{\hbox{$^{\hat{f}}$\hspace{-0.15cm}
$\mathcal{D}$}}

\newcommand{\T}{{\rm tr}}

\newcommand{\bb}{\begin{eqnarray}}
\newcommand{\ee}{\end{eqnarray}}
\newcommand{\eee}{\nonumber\end{eqnarray}}
\newcommand{\pp}[1]{\begin{pmatrix} #1 \end{pmatrix}}

\newcommand{\rxyd}[2]{{\begin{xy} 0;<2mm,0mm>:<0mm,2mm>::0;0,
,(5,-2)*{a}
,(10,-1.8)*{\bar{a}}
,(15,-2)*{b}
,(20,-1.8)*{\bar{b}}
,(25,-2)*{c}
,(2,-5)*{a}
,(2,-10)*{\bar{a}}
,(2,-15)*{b}
,(2,-20)*{\bar{b}}
,(2,-25)*{c}
,(5,-5)*\cir(#1,0){}
,(10,-5)*\cir(#1,0){}
,(15,-5)*\cir(#1,0){}
,(20,-5)*\cir(#1,0){}
,(25,-5)*\cir(#1,0){}
,(5,-10)*\cir(#1,0){}
,(10,-10)*\cir(#1,0){}
,(15,-10)*\cir(#1,0){}
,(20,-10)*\cir(#1,0){}
,(25,-10)*\cir(#1,0){}
,(5,-15)*\cir(#1,0){}
,(10,-15)*\cir(#1,0){}
,(15,-15)*\cir(#1,0){}
,(20,-15)*\cir(#1,0){}
,(25,-15)*\cir(#1,0){}
,(5,-20)*\cir(#1,0){}
,(10,-20)*\cir(#1,0){}
,(15,-20)*\cir(#1,0){}
,(20,-20)*\cir(#1,0){}
,(25,-20)*\cir(#1,0){}
,(5,-25)*\cir(#1,0){}
,(10,-25)*\cir(#1,0){}
,(15,-25)*\cir(#1,0){}
,(20,-25)*\cir(#1,0){}
,(25,-25)*\cir(#1,0){}
#2\end{xy}}}

\newcommand{\rxydd}[2]{{\begin{xy} 0;<2mm,0mm>:<0mm,2mm>::0;0,
,(5,-2)*{a}
,(10,-1.8)*{\bar{a}}
,(15,-2)*{b}
,(20,-1.8)*{c}
,(25,-2)*{\bar{c}}
,(2,-5)*{a}
,(2,-10)*{\bar{a}}
,(2,-15)*{b}
,(2,-20)*{c}
,(2,-25)*{\bar{c}}
,(5,-5)*\cir(#1,0){}
,(10,-5)*\cir(#1,0){}
,(15,-5)*\cir(#1,0){}
,(20,-5)*\cir(#1,0){}
,(25,-5)*\cir(#1,0){}
,(5,-10)*\cir(#1,0){}
,(10,-10)*\cir(#1,0){}
,(15,-10)*\cir(#1,0){}
,(20,-10)*\cir(#1,0){}
,(25,-10)*\cir(#1,0){}
,(5,-15)*\cir(#1,0){}
,(10,-15)*\cir(#1,0){}
,(15,-15)*\cir(#1,0){}
,(20,-15)*\cir(#1,0){}
,(25,-15)*\cir(#1,0){}
,(5,-20)*\cir(#1,0){}
,(10,-20)*\cir(#1,0){}
,(15,-20)*\cir(#1,0){}
,(20,-20)*\cir(#1,0){}
,(25,-20)*\cir(#1,0){}
,(5,-25)*\cir(#1,0){}
,(10,-25)*\cir(#1,0){}
,(15,-25)*\cir(#1,0){}
,(20,-25)*\cir(#1,0){}
,(25,-25)*\cir(#1,0){}
#2\end{xy}}}

\newcommand{\rxyddd}[2]{{\begin{xy} 0;<2mm,0mm>:<0mm,2mm>::0;0,
,(5,-2)*{a}
,(10,-1.8)*{\bar{a}}
,(15,-2)*{b}
,(20,-1.8)*{\bar{b}}
,(25,-2)*{c}
,(30,-1.8)*{\bar{c}}
,(2,-5)*{a}
,(2,-10)*{\bar{a}}
,(2,-15)*{b}
,(2,-20)*{\bar{b}}
,(2,-25)*{c}
,(2,-30)*{\bar{c}}
,(5,-5)*\cir(#1,0){}
,(10,-5)*\cir(#1,0){}
,(15,-5)*\cir(#1,0){}
,(20,-5)*\cir(#1,0){}
,(25,-5)*\cir(#1,0){}
,(30,-5)*\cir(#1,0){}
,(5,-10)*\cir(#1,0){}
,(10,-10)*\cir(#1,0){}
,(15,-10)*\cir(#1,0){}
,(20,-10)*\cir(#1,0){}
,(25,-10)*\cir(#1,0){}
,(30,-10)*\cir(#1,0){}
,(5,-15)*\cir(#1,0){}
,(10,-15)*\cir(#1,0){}
,(15,-15)*\cir(#1,0){}
,(20,-15)*\cir(#1,0){}
,(25,-15)*\cir(#1,0){}
,(30,-15)*\cir(#1,0){}
,(5,-20)*\cir(#1,0){}
,(10,-20)*\cir(#1,0){}
,(15,-20)*\cir(#1,0){}
,(20,-20)*\cir(#1,0){}
,(25,-20)*\cir(#1,0){}
,(30,-20)*\cir(#1,0){}
,(5,-25)*\cir(#1,0){}
,(10,-25)*\cir(#1,0){}
,(15,-25)*\cir(#1,0){}
,(20,-25)*\cir(#1,0){}
,(25,-25)*\cir(#1,0){}
,(30,-25)*\cir(#1,0){}
,(5,-30)*\cir(#1,0){}
,(10,-30)*\cir(#1,0){}
,(15,-30)*\cir(#1,0){}
,(20,-30)*\cir(#1,0){}
,(25,-30)*\cir(#1,0){}
,(30,-30)*\cir(#1,0){}
#2\end{xy}}}

\newcommand{\rxyzz}[2]{{\begin{xy} 0;<2mm,0mm>:<0mm,2mm>::0;0,
,(5,-2)*{a} ,(10,-2)*{b} ,(15,-1.8)*{\bar{b}} ,(2,-5)*{a} ,(2,-10)*{b}
,(2,-14.8)*{\bar{b}} ,(5,-5)*\cir(#1,0){} ,(10,-5)*\cir(#1,0){}
,(15,-5)*\cir(#1,0){}
,(5,-10)*\cir(#1,0){} ,(10,-10)*\cir(#1,0){} ,(15,-10)*\cir(#1,0){}
,(5,-15)*\cir(#1,0){} ,(10,-15)*\cir(#1,0){} ,(15,-15)*\cir(#1,0){}
#2\end{xy}}}

\begin{document}

\thispagestyle{empty}

\begin{center}
CENTRE DE PHYSIQUE TH\'EORIQUE \footnote{\, Unit\'e Mixed de
Recherche (UMR) 6207 du CNRS et des Universit\'es Aix-Marseille 1 et 2 \\ \indent \quad \, Sud Toulon-Var, Laboratoire affili\'e \`a la 
FRUMAM (FR 2291)} \\ CNRS--Luminy, Case 907\\ 13288 Marseille Cedex 9\\
FRANCE
\end{center}

\vspace{1.5cm}

\begin{center}
{\Large\textbf{On a Classification of Irreducible \\ Almost Commutative
Geometries, \\[0.3cm] A Second Helping}} 
\end{center}

\vspace{1.5cm}

\begin{center}
{\large Jan-Hendrik Jureit
\footnote{\, and Universit\'e de Provence and Universit\"at Kiel}${}^,$
\footnote{\, jureit@cpt.univ-mrs.fr} 
\quad Christoph A. Stephan$\,\,^{2,}$
\footnote{\, stephan@cpt.univ-mrs.fr} }

\vspace{1.5cm}

{\large\textbf{Abstract}}
\end{center}

We complete the classification of almost commutative geometries 
from a particle physics point of view given in \cite{class}. Four 
missing Krajewski diagrams will be presented after a short introduction
into irreducible, non-degenerate spectral triples.

\vspace{1.5cm}

\vskip 1truecm

PACS-92: 11.15 Gauge field theories\\
\indent MSC-91: 81T13 Yang-Mills and other gauge theories

\vskip 1truecm

\noindent January 2005
\vskip 1truecm
\noindent CPT-2005/P.002\\
\noindent

\newpage

\section{Introduction}

Alain Connes' noncommutative geometry \cite{book,real,grav,cc} allows in an elegant way
to unify gravity and the standard model of particle physics. 
A central role in this formalism
is played by almost commutative spectral triples ($\mathcal{A},\mathcal{H},\mathcal{D}$) which decompose into an external and an internal, finite dimensional component. The external
part encodes a compact 4-dimensional Euclidian spacetime and the internal one corresponds to 
a discrete 0-dimensional Kaluza-Klein space, determining the particle content of the theory.
Via the spectral action  \cite{cl} one recovers the Einstein-Hilbert action combined with the bosonic
action of a Yang-Mills-Higgs (YMH) theory. Since the set of allowed YMH-theories is determined
by the possible internal, finite dimensional spectral triples, we will restrict ourselves
to this part. The standard model of particle physics is the most prominent example in this context.

Real, finite dimensional spectral triples have been completely classified by Krajewski \cite{Kraj} and Paschke \& Sitarz \cite{pasch}. A classification of almost commutative geometries from a physical point of view
was given in \cite{class}. The spectral triples were required to be irreducible and non-degenerate, in the sense that the 
Hilbert space was chosen to be as small as possible with non-degenerate fermion masses.
Heavy use was made of Krajewski's diagrammatic method, which will be described briefly below.  
The main obstacle in finding all physically relevant almost
commutative spectral triples is the sheer mass of diagrams which have to be considered. Since this is 
a purely combinatorial problem it is convenient to let a computer do the tedious task. The cases of one
and two matrix algebras can still be done by hand. But already three
algebras produce hundreds of diagrams and one easily looses sight.

Therefore we developed an algorithm to calculate these diagrams and used the known results 
from \cite{class} to test and calibrate the program. The main goal was to extend the calculations
to more than three algebras, where we expect thousands of possible irreducible spectral triples. 
During the calibration it turned out that four diagrams were overlooked in the case of three algebras. To 
complete the proof we will present these four missing diagrams and their models in this paper. 
The algorithm to compute the diagrams and the results for the case of four algebras will be presented 
elsewhere.

In section 2 and 3 we will briefly introduce Krajewski diagrams and irreducible, 
non-degenerate spectral triples.
The missing diagrams for three algebras will be presented in section 4. 

\section{Basic definitions and Krajewski diagrams}
In this section we will give the necessary basic definitions for a classification of almost commutative
geometries from a particle physics point of view. As mentioned above only the 0-dimensional part
will be taken into account, so we restrict ourselves to real, $S^0$-real, finite spectral triples
($\mathcal{A},\mathcal{H},\mathcal{D}, $ $J,\epsilon,\chi$). The algebra $\mathcal{A}$ is
a finite sum of matrix algebras
$\mathcal{A}= \oplus_{i=1}^{N} M_{n_i}(\mathbb{K}_i)$ with $\mathbb{K}_i=\mathbb{R},\mathbb{C},\mathbb{H}$ where $\mathbb{H}$
denotes the quaternions. 
A faithful representation $\rho$ of $\mathcal{A}$ is given on the finite dimensional Hilbert space $\mathcal{H}$.
The Dirac operator $\mathcal{D}$ is a selfadjoint operator on $\mathcal{H}$ and plays the role of the fermionic mass matrix.
$J$ is an antiunitary involution, $J^2=1$, and is interpreted as the charge conjugation
operator of particle physics.
The $S^0$-real structure $\epsilon$ is a unitary involution, $\epsilon^2=1$. Its eigenstates with
eigenvalue $+1$ are the particle states, eigenvalue $-1$ indicates antiparticle states. 
The chirality $\chi$ is as well a unitary involution, $\chi^2=1$, whose eigenstates with eigenvalue
$+1$ $(-1)$ are interpreted as right (left) particle states.
These operators are required to fulfill Connes' axioms for spectral triples:

\begin{itemize}
\item  $[J,\mathcal{D}]=[J,\chi]=[\epsilon,\chi]=[\epsilon,\mathcal{D}]=0, \quad \epsilon
J=-J \epsilon,\quad\mathcal{D}\chi =-\chi \mathcal{D}$, 
 
$[\chi,\rho(a)]=[\epsilon,\rho(a)]=[\rho(a),J\rho(b)J^{-1}]=
[[\mathcal{D},\rho(a)],J\rho(b)J^{-1}]=0, \forall a,b \in \mathcal{A}$.
\item The chirality can be written as a finite sum $\chi =\sum_i\rho(a_i)J\rho(b_i)J^{-1}.$
This condition is called {\it orientability}.
\item The intersection form
$\cap_{ij}:=\T(\chi \,\rho (p_i) J \rho (p_j) J^{-1})$ is non-degenerate,
$\rm{det}\,\cap\not=0$. The
$p_i$ are minimal rank projections in $\mathcal{A}$. This condition is called
{\it Poincar\'e duality}.
\end{itemize} 
Now the Hilbert space $\mathcal{H}$ and the representation $\rho$ decompose with respect to the 
eigenvalues of $\epsilon$ and $\chi$ into left and right, particle and antiparticle spinors
and representations:
\begin{eqnarray}
\mathcal{H}=\mathcal{H}_L\oplus\mathcal{H}_R\oplus\mathcal{H}_L^c\oplus\mathcal{H}_R^c 
\label{hilbertspace}
\end{eqnarray}
\begin{eqnarray}
\rho = \rho_L \oplus \rho_R \oplus \overline{ \rho_L^c} \oplus \overline{ \rho_R^c}
\label{representation}
\end{eqnarray}
In this representation the Dirac operator has the form
\begin{eqnarray}
\mathcal{D}=\pp{0&\mathcal{M}&0&0\\
\mathcal{M}^*&0&0&0\\ 0&0&0&\overline{\mathcal{M}}\\
0&0&\overline{\mathcal{M}^*}&0}, \label{opdirac}
\end{eqnarray}
where $\mathcal{M}$ is the fermionic mass matrix connecting the left and the right handed Fermions.

Since the individual matrix algebras have only one fundamental representation for $\mathbb{K}=
\mathbb{R},\mathbb{H}$ and two for $\mathbb{K}=\mathbb{C}$ (the fundamental one and its complex
conjugate), $\rho$ may be written as a direct sum of these fundamental representations with 
mulitiplicities
\begin{eqnarray}
\rho(\oplus_{i=1}^N a_i):=(\oplus_{i,j=1;\alpha_i,\alpha_j}^N
a_{i\alpha_{i}} \otimes
1_{m_{j\alpha_{j}i\alpha_{i}}} \otimes 1_{(n_j)})\
\oplus\ ( \oplus_{i,j=1}^N 1_{(n_i)} \otimes 1_{m_{j\alpha_ji\alpha_i}} \otimes
\overline{a_{j\alpha_{j}}} ).
\label{kranotcomplex}
\end{eqnarray}
The first summand denotes the particle sector and the second the antiparticle sector. For the dimensions
of the unity matrices we have $(n)=n$ for $\mathbb{K}=\mathbb{R},\mathbb{C}$ and $(n)=2n$ for
$\mathbb{K}=\mathbb{H}$ and the convention $1_0=0$. The index $\alpha$ indicates wether the representation 
is the fundamental one or its complex conjugate and thus applies only to the $\mathbb{K}=\mathbb{C}$ case.
The multiplicities $m_{j\alpha_ji\alpha_i}$ are non-negative integers. Acting with the real structure
$J$ on $\rho$ permutes the main summands and complex conjugates them. It is also possible to write
the chirality as a direct sum 
\begin{eqnarray}
\chi=(\oplus_{i,j=1}^N 1_{(n_i)} \otimes \chi_{j\alpha_ji\alpha_i}1_{m_{j\alpha_ji\alpha_i}} \otimes
1_{(n_j)})\
\oplus\
(\oplus_{i,j=1}^N 1_{(n_i)} \otimes \chi_{j\alpha_ji\alpha_i}1_{m_{j\alpha_ji\alpha_i}} \otimes 1_{(n_j)}),
\end{eqnarray}
where $\chi_{j\alpha_ji\alpha_i}=\pm 1$ according to our previous convention on left-(right-)handed spinors.
One can now define the multiplicity matrix $\mu \in M_N(\mathbb{Z})$ such that 
$\mu _{j\alpha_ji\alpha_i}:=\chi _ {j\alpha_ji\alpha_i}\, m_{j\alpha_ji\alpha_i}$. This matrix is symmetric and decomposes into a particle and an antiparticle matrix, the second being just the particle matrix transposed, $\mu= \mu_P + \mu_A = \mu_P + \mu_P^T$. The entries of the multiplicity matrix must fulfill
certain consistency conditions, which are given by table 1 in \cite{Kraj}.
It is also possible to recover the intersection form of the Poincar\'e duality, up to a numerical factor,
if the multiplicity matrix is contracted by summation over the $\alpha$'s.

The mass matrix $\mathcal{M}$ of the Dirac operator connects the left and the right handed Fermions. Using
the decomposition of the representation $\rho$ and the corresponding decomposition of the Hilbert
space $\mathcal{H}$ we find two types of submatrices in $\mathcal{M}$, namely $M\otimes 1_{(n_k)}$ and
$1_{(n_k)}\otimes M$. $M$ is a complex $(n_i)\times(n_j)$ matrix connecting the i-th and the j-th sub-Hilbert
space and its eigenvalues give the masses of the fermion multiplet. We will call the k-th algebra the
colour algebra.

Connes' axioms, the decomposition of the Hilbert space, the representation and the Dirac operator
allow a diagrammatic dipiction. As was shown in \cite{Kraj} and \cite{class} this can be boiled down
to simple arrows, which encode the multiplicity matrix and the fermionic mass matrix.
From these information all the ingredients of the spectral triple can be recovered.
For our purpose a simple arrow and a double arrow are sufficient. The arrows allways point from
right fermions (positive chirality) to left fermions (negative chirality).
 We may also restrict ourselves to
the particle part, since the information of the antiparticle part is included by transposing the
particle part. As an example we will retranslate these two generic arrows 
into the multiplicity matrix, the representation of the algebra and the Dirac operator. We will adopt
the conventions of \cite{class} so that algebra elements tensorised with $1_{m_{ij}}$ will be written
as a direct sum of $m_{ij}$ summands.

Take the algebra
$\mathcal{A}=\mathbb{H}\oplus M_3(\mathbb{C})\owns (a,b)$ with the first diagram of Figure 1.
\begin{center}
\begin{tabular}{ccc}
\rxyzz{0.7}{
,(5,-5);(10,-5)**\dir{-}?(0.4)*\dir{<}
}
&
\;\;\;\;\;
\rxyzz{0.7}{
,(5,-5)*\cir(0.4,0){}*\frm{*}
,(5,-5);(10,-5)**\dir2{-}?(0.4)*\dir2{<}
}
\\
\\ 
& \hspace{-2cm} Fig. 1&
\end{tabular}
\end{center}
Then the multiplicity matrix and its contraction are
\bb
\mu=\pp{-1&1&0 \cr 0&0&0 \cr 0&0&0},\quad
\hat\mu=\pp{-1&1 \cr 0&0 }.
\eee
Using (\ref{representation}), its representation is, up to
unitary equivalence
\bb
\rho_L (a,b)=a \otimes 1_2,\, \rho _R (a,b) =b \otimes 1_2,\, \rho_L^c (a,b)=1_2\otimes
a,\,
\rho _R^c(a,b)=1_3\otimes a.
\eee The Hilbert space is
\bb
\mathcal{H}=\mathbb{C}^4 \oplus \mathbb{C}^6 \oplus \mathbb{C} ^4 \oplus \mathbb{C}^6.
\eee
In its Dirac operator (\ref{opdirac}) the mass matrix is $\mathcal{M}=M\otimes 1_2$,
where $M$ is a nonvanishing complex $2\times 3 $
matrix.

\noindent Real structure, $S^0$-real structure and chirality are given by
(cc stands for
complex conjugation)
\bb J=
\begin{pmatrix} 0&1_{10}\\ 1_{10}&0
\end{pmatrix}
\circ {\rm cc},\quad
\epsilon =
\begin{pmatrix} 1_{10}&0\\ 0&-1_{10}
\end{pmatrix} ,\quad \chi =
\begin{pmatrix}-1_4&0&0&0\\ 0&1_6&0&0\\ 0&0&-1_4&0\\ 0&0&0&1_6
\end{pmatrix} .\eee The first tensor factor in $a\otimes 1_2$ concerns
particles, the second concerns antiparticles denoted by
$\cdot^c$.  The antiparticle representation is read from the transposed
multiplicity matrix.

The second diagram of Figure 1 yields
\bb
\mu=\pp{-1&2&0 \cr 0&0&0 \cr 0&0&0},\quad
\hat\mu=\pp{-1&2 \cr 0&0},
\eee
and its spectral triple reads:
\bb
\rho _L(a,b)=a\otimes 1_2,\quad \rho _R(a,b)=\pp{b&0\\ 0& b}\otimes 1_2,
\nonumber\\[2mm]
\rho _L^c(a,b)=1_2\otimes a,\quad
\rho _R^c(a,b)=\pp{1_3&0\\ 0&1_3}\otimes a, \cr \cr
\mathcal{M}=\pp{M_1&M_2}\otimes 1_2,\, M_1\ {\rm and}\ M_2
\ {\rm of \ size}\ 2\times 3,
\eee
\vspace{-0.9cm}
\bb
 J=\pp{0&1_{16}\\ 1_{16}&0}\circ {\rm cc},\quad
\epsilon =\pp{1_{16}&0\\ 0&-1_{16}},\quad \chi =\pp{-1_4&0&0&0\\
0&1_{12}&0&0\\ 0&0&-1_4&0\\ 0&0&0&1_{12}}.
\eee

It should be clear that the number of possible ways to fit one and more arrows into a diagram increases
factorially with the number of matrix algebras, i.e. with the size of the diagram. Thus this seemingly
simple problem soon becomes intractable, even for a powerful computer.

We started out with the flat Dirac operator of a 4-dimensional spacetime with a fixed fermionic
mass matrix. To generate curvature we have to perform a general coordinate transformation and then
fluctuate the Dirac operator. This can be achieved by lifting the automorphisms of the algebra to
the Hilbert space, unitarily transforming the Dirac operator with the lifted automorphisms
and then building linear combinations. Again we restrict ourselves to the finite case.
Except for complex conjugation in $M_n(\mathbb{C})$ and permutations of
identical summands in the algebra $\mathcal{A}=\mathcal{A}_1\oplus\mathcal{A}_2\oplus ...\oplus\mathcal{A}_N$,
every algebra automorphism
$\sigma
$  is inner, $\sigma (a)=uau^{-1}$ for a unitary $ u\in U(\mathcal{A})$. Therefore
the connected component of the automorphism group is
Aut$(\mathcal{A})^e=U(\mathcal{A})/(U(\mathcal{A})\cap{\rm Center}(\mathcal{A}))$. Its lift to the Hilbert
space \cite{real}
\bb L(\sigma )=\rho (u)J\rho (u)J^{-1}\eee is multi-valued.

The {\it fluctuation $\ddf$} of the Dirac operator $\mathcal{D}$ is given by a
finite collection $f$ of real numbers
$r_j$ and algebra automorphisms $\sigma _j\in{\rm Aut}(\mathcal{A})^e$ such
that
\bb
\ddf :=\sum_j r_j\,L(\sigma _j) \, \mathcal{D} \, L(\sigma_j)^{-1},\quad r_j\in\mathbb{R},\
\sigma _j\in{\rm Aut}(\mathcal{A})^e.
\eee
The fluctuated Dirac operator $\ddf$ is often denoted by $\varphi $, the
`Higgs scalar', in the physics literature.  We consider only fluctuations
with real coefficients since
$\ddf$ must remain selfadjoint.

To avoid the multi-valuedness in the fluctuations, we allow the entire
unitary group viewed as a (maximal) central extension of the
automorphism group.

An almost commutative geometry is the tensor product of a finite
noncommutative triple with an infinite, commutative spectral triple. By
Connes' reconstruction theorem \cite{grav} we know that the latter comes
from a Riemannian spin manifold, which we will take to be any
4-dimensional, compact, flat manifold like the flat 4-torus.  The spectral
action of this almost commutative spectral triple reduced to the finite part
is a functional on the vector space of all fluctuated, finite Dirac operators:
\bb V(\ddf )= \lambda\  \T\!\left[ (\ddf )^4\right] -\textstyle{\frac{\mu
^2}{2}}\
\T\!\left[
(\ddf) ^2\right] ,\eee where $\lambda $ and $\mu $ are positive constants
\cite{cc,kraj2}.
The spectral action is invariant under lifted automorphisms and even
under the unitary group $U(\mathcal{A})\owns u$,
\bb V( [\rho (u)J\rho (u)J^{-1}] \, \ddf \, [\rho (u)J\rho
(u)J^{-1}]^{-1})=V(\ddf),\eee and it is bounded from below.
 Our task is to find the minima $  \ddfm $ of this action and
their spectra.

\section{Irreducibility, Non-Degeneracy}

To classify the almost commutative spectral triples we will impose some extra conditions
as in \cite{class}. We will require the spectral triples to be irreducible and non-degenrate
according to the following definitions:

\begin{defn}
 i) A spectral triple $(\mathcal{A},\mathcal{H},\mathcal{D})$ is {\it degenerate} if the kernel of
$\mathcal{D}$ contains a non-trivial subspace of the complex Hilbert space $\mathcal{H}$
invariant under the representation $\rho$ on $\mathcal{H}$ of the real algebra
 $\mathcal{A}$.  \\
 ii) A non-degenerate spectral triple $(\mathcal{A},\mathcal{H},\mathcal{D})$ is {\it reducible} if
there is a proper subspace
$\mathcal{H}_0\subset\mathcal{H}$ invariant under the algebra $\rho(\mathcal{A})$  such that
$(\mathcal{A},\mathcal{H}_0,\mathcal{D}|_{\mathcal{H}_0})$ is a non-degenerate spectral triple. If the
triple is real, $S^0$-real and even, we require  the subspace
$\mathcal{H}_0$ to be also invariant under the real structure $J$, the $ S^0$-real
structure $\epsilon $ and under the chirality
$\chi $ such that the triple $(\mathcal{A},\mathcal{H}_0,\mathcal{D}|_{\mathcal{H}_0})$ is again real,
$S^0$-real and even.
\end{defn}

\begin{defn} The irreducible
 spectral triple $(\mathcal{A},\mathcal{H},\mathcal{D})$ is {\it dynamically non-degenerate} if all
minima $\ddfm$ of the action $V(\ddf)$ define a non-degenerate spectral
triple $(\mathcal{A},\mathcal{H},\ddfm )$ and if the spectra of all minima  have no
degeneracies other than the three kinematical degeneracies: left-right,
particle-antiparticle and colour. Of course in the massless case there is no
left-right degeneracy. We also suppose that the colour degeneracies are
protected by the little group. By this we mean that all eigenvectors of
$\ddfm$ corresponding to the same eigenvalue are in a common orbit of
the little group (and scalar multiplication and charge conjugation).
\label{irred}
\end{defn}

In physicists' language non-degeneracy excludes all models with pairwise equal fermion masses
in the left handed particle sector up to colour degenracy. 
Irreducibility tells us that the Krajewski diagrams, which we have to find
must not contain more arrows than strictly necessary to satisfy Connes' Axioms, especially the Poincar\'e
duality. The last requirement of definition \ref{irred} means noncommutative colour groups are unbroken. It ensures that the corresponding mass degeneracies are protected from quantum corrections.

\section{The Missing Diagrams}
In this section we will present the diagrams missing in the proof for three algebras in \cite{class}.
For every diagram only one representative model will be given. All the other models can be obtained
by simply exchanging left with right and particles with antiparticles.
On the diagrammatic side
this is equivalent to changing the directions of all arrows or reflecting the diagram on its diagonal. 
Permutations of the diagrams are neglected as well, since they lead to the same physical models with
a different order of the particles. For every diagram there are several ways to connect the algebras 
by arrows in accordance with the consistency conditions of table 1. in \cite{Kraj}. 
With respect to this, the four diagrams are all computed in the same way and they all fall in the same way.
The possibilities of complex conjugating an algebra representation are limited and yield no essentially
new models. It should be obvious from the diagrams wether the matrix algebras are complex, real or 
quaternionic. In all other cases the choice of the field will not affect the calculations, so we 
will not specify the algebras explicitly.
For the four missing diagrams, 
$\mathcal{A}_1,\mathcal{A}_2,\mathcal{A}_3$ denote the algebras, $a,b,c$ their generic elements and
$k,\ell,p$ the respective size of the matrices.

\vspace{1\baselineskip}
{\bf Diagram 1} yields the representation
\bb
\rho _L (a,b,c) =\pp{b\otimes 1_k&0&0\\ 0&c\otimes 1_k&0\\ 0& 0&b\otimes 1_p}, \,
\rho _R (a,b,c) =\pp{\bar a \otimes 1_k&0&0\\ 0&\bar b \otimes 1_k&0\\
 0&0&a\otimes 1_p},\cr \cr \cr
\rho _L^c (a,b,c) =\pp{1_\ell\otimes a&0&0\\ 0&1_p\otimes a&0\\ 0& 0&1_\ell\otimes \bar c}, \,
\rho _R^c (a,b,c) =\pp{1_k\otimes a&0&0\\ 0&1_\ell\otimes a&0\\ 0& 0&1_p\otimes \bar c}.
\eee  The mass matrix is
\bb \mathcal{M}=\pp{M_1\otimes 1_k&0&0\\ 0&M_2\otimes 1_k &0\\ 0&0&M_3\otimes
1_p},\, M_1,M_3\in M_{k\times\ell}(\mathbb{C}),\,  M_2\in M_{p\times \ell}(\mathbb{C}),
\eee
where all three algebras are $M_n(\mathbb{C})$.
The fluctuations are
\bb
\Mf_1  &=&\sum_j r_j\,v _jM_1 \bar u_j^{-1},\quad u_j\in U(\mathcal{A}_1),\quad v_j\in
U(\mathcal{A}_2),\cr
\Mf_2  &=&\sum_j r_j\,w _jM_2 \bar v_j^{-1},\quad w_j\in U(\mathcal{A}_3),\cr
\Mf_3  &=&\sum_j r_j\,v _jM_3 u_j^{-1},\eee and the action
$V(C_1,C_2,C_3)$ is, with
$C_i:={\Mf_i}^* \, \Mf_i$ equal to
\bb 4k \, [\lambda\, {\rm tr} (C_1)^2 -  {\textstyle\frac{1}{2}} \mu ^2\,
{\rm tr} (C_1)] +4k \, [\lambda\, {\rm tr} (C_2)^2- {\textstyle\frac{1}{2}} \mu
^2 \,{\rm tr} (C_2)] +4p \, [\lambda\, {\rm tr} (C_3)^2-
{\textstyle\frac{1}{2}}
\mu ^2 \,{\rm tr} (C_3)].
\eee
Counting neutrinos and imposing broken colour to be commutative leaves
only one case, $k=\ell=p=1$. The fluctuations decouple the $\Mf_i$ so it is 
allways possible to reach the absolute minimum of the Higgs potential and the triple
is degenerate.

\vspace{1\baselineskip}

{\bf Diagram 2} falls in the same way.

\vspace{1\baselineskip}

{\bf Diagram 3} is degenerate in the commutative case and exhibits mass relations
in the noncommutative case. The calculation runs along the lines of diagram 8 in
\cite{class}.
\vspace{1\baselineskip} 

{\bf Diagram 4} yields the representation
\bb \rho _L (a,b,c) =\pp{c\otimes 1_k&0\\ 0&a\otimes 1_\ell},&&
\rho _R (a,b,c) =\pp{\bar b\otimes 1_k&0&0\\ 0&\bar b\otimes 1_k&0 \\ 0&0&c\otimes 1_\ell},\cr \cr \cr
\rho _L^c (a,b,c) =\pp{1_p\otimes a&0\\ 0&1_k\otimes b},&&
\rho _R^c (a,b,c) =\pp{1_\ell\otimes a&0&0\\ 0&1_\ell\otimes a&0 \\ 0&0& 1_p\otimes b},\eee with possible
complex conjugations here and there. The mass matrix is
\bb \mathcal{M}=\pp{M_1\otimes 1_k& M_2\otimes 1_k&0\\ 0&0&M_3\otimes 1_\ell},\quad M_1,
M_2\in M_{p\times\ell}(\mathbb{C}), \quad M_3\in M_{k\times p}(\mathbb{C}).\eee The fluctuations are
\bb
\Mf_1  &=&\sum_j r_j\,w _jM_1 \bar v_j^{-1},\quad w_j\in U(\mathcal{A}_3),\quad v_j\in
U(\mathcal{A}_2),\cr
\Mf_2  &=&\sum_j r_j\,w _jM_2 \bar v_j^{-1},\cr
\Mf_3  &=&\sum_j r_j\,u _jM_3 w_j^{-1},\quad u_j\in U(\mathcal{A}_1)
\eee and the action is
\bb V(C_1,C_2,C_3) = 4k \, [\lambda\, {\rm tr} (C_1+C_2)^2 -
{\textstyle\frac{1}{2}} \mu ^2\, {\rm tr} (C_1+C_2)] + 4p \, [\lambda\,
{\rm tr}
(C_3)^2 - {\textstyle\frac{1}{2}} \mu ^2 \,{\rm tr} (C_3)].
\eee The neutrino count and broken colour imply $k=\ell=1$ and $p=1$ or $p=2$.

\noindent The case $k=\ell =p=1$ is obviously degenerate.

\vspace{1\baselineskip} 

For $k=\ell=1$, $p=2$ we have one neutrino. $\Mf_3$ fluctuates independently and can
be pushed into the absolute minimum of the Higgs potential.
Let us put $\Mf_1$ and $\Mf_2$ into one matrix
\bb
\Mf_{1,2} = \sum_j r_j\,w _j (M_1,M_2) \pp{\bar v_j^{-1} &0 \\ 0&\bar v_j^{-1}}
\eee
Since the $\bar v_j^{-1} \in \mathbb{C}$ they commute with $(M_1,M_2)$ and so 
\bb
\Mf_{1,2} = C (M_1,M_2),
\eee
where $C\in M_{2\times 2}(\mathbb{C})$ is an arbitrary matrix. $M_1$ has to be lineary
independent of $M_2$ because otherwise they would produce a second neutrino. It follows
that $(M_1,M_2)$ is invertible and we can choose $C$ to be its inverse. In this way
we reach the absolute minimum of the Higgs potential and the triple is degenerate.

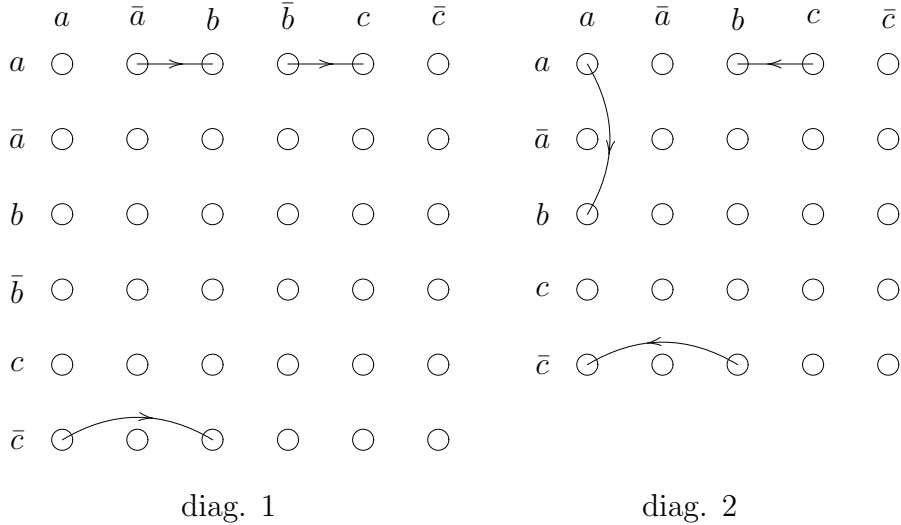
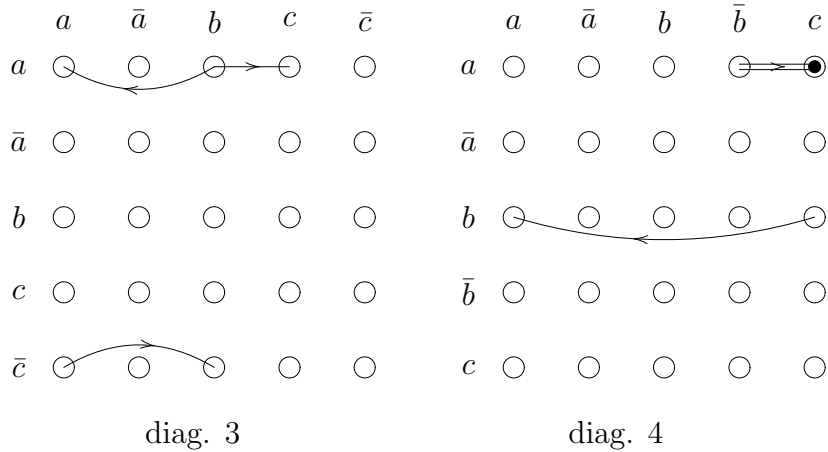
\begin{center}
\begin{tabular}{cc}
\rxyddd{0.7}{
,(10,-5);(15,-5)**\dir{-}?(.6)*\dir{>}
,(20,-5);(25,-5)**\dir{-}?(.6)*\dir{>}
,(5,-30);(15,-30)**\crv{(10,-27)}?(.6)*\dir{>}
}
&
\;\;\;\;\;
\rxydd{0.7}{
,(5,-5);(5,-15)**\crv{(8,-10)}?(.6)*\dir{>}
,(20,-5);(15,-5)**\dir{-}?(.6)*\dir{>}
,(15,-25);(5,-25)**\crv{(10,-22)}?(.6)*\dir{>}
}

\\
\\ 
diag. 1 & diag. 2 
\end{tabular}
\end{center}
\begin{center}
\begin{tabular}{cc}
\rxydd{0.7}{
,(5,-5);(15,-5)**\crv{(10,-8)}?(.4)*\dir{<}
,(20,-5);(15,-5)**\dir{-}?(.4)*\dir{<}
,(15,-25);(5,-25)**\crv{(10,-22)}?(.4)*\dir{<}
}
&
\;\;\;\;\;

\rxyd{0.7}{
,(25,-5)*\cir(0.4,0){}*\frm{*}
,(20,-5);(25,-5)**\dir2{-}?(.6)*\dir{>}
,(25,-15);(5,-15)**\crv{(15,-18)}?(.6)*\dir{>}
}
\\
\\
diag. 3 & diag. 4 
\end{tabular}
\end{center}

\section{Conclusion} 
The new models discovered with help of the computer complete the proof for up to 
three algebras given in \cite{class}. We did not find anything of interest from the
particle physics point of view but we gained confidence in our algorithm and it seems
sensible to compute the case with four algebras. 

\vskip1cm
\noindent
{\bf Acknowledgements:} The authors would like to thank T. Sch\"ucker for his advice and support. We
gratefully acknowledge our fellowships of the Friedrich-Ebert-Stiftung.


\begin{thebibliography}{10}

\bibitem{class}
B. Iochum, T. Sch\"ucker, C. Stephan,
``On a Classification of Irreducible Almost Commutative Geometries'',
hep-th/0312276, J. Math. Phys. in press
\bibitem{cc}
 A. Chamseddine \& A. Connes, ``The spectral action principle'',
hep-th/9606001, Comm. Math. Phys. 182 (1996) 155
\bibitem{book}
 A. Connes, {\it Noncommutative Geometry}, Academic Press, London and San
Diego (1994)
\bibitem{real}
A. Connes, ``Noncommutative geometry and reality'',  J.
Math. Phys. 36 (1995) 6194
\bibitem{grav}
A. Connes, `` Gravity coupled with matter and the
foundation of noncommutative geometry'', hep-th/9603053, Comm. Math.
Phys. 155 (1996) 109
\bibitem{cl}
A. Connes \& J. Lott, ``Particle models and noncommutative
geometry'', Nucl. Phys. B 18B (1990) 29\\
 A. Connes \& J. Lott,``The metric
aspect of noncommutative geometry'', in the
proceedings of the 1991 Carg\`ese Summer Conference,
eds.: J. Fr\"ohlich et al., Plenum Press (1992)
\bibitem{Kraj}
T. Krajewski, ``Classification of finite spectral triples'',
hep-th/9701081, J. Geom. Phys. 28 (1998) 1
\bibitem{pasch}
M. Paschke \& A. Sitarz, ``Discrete spectral triples and
their symmetries'', q-alg/9612029, J. Math. Phys. 39 (1998) 6191
\bibitem{kraj2}
T. Krajewski, ``Constraints on scalar potential from spectral action
principle'', hep-th/9803199
\end{thebibliography}
\end{document}